\documentclass[twocolumn]{aastex631}

\usepackage[usenames,dvipsnames]{xcolor}

\begin{document}

\title{To What Extent Are Star Cluster Ages Encoded in Their Environments? Exploring the Spatial Distribution of Age-Related Information with PHANGS-HST Imaging and Convolutional Neural Networks}

\author[0000-0002-0563-784X]{Javier Via\~na}
\affiliation{Department of Physics and Kavli Institute for Astrophysics and Space Research, Massachusetts Institute of Technology, Cambridge, MA 02139, USA}

\author[0000-0003-0946-6176]{Janice C. Lee}
\affiliation{Space Telescope Science Institute, 3700 San Martin Drive, Baltimore, MD 21218, USA}

\author[0000-0001-7246-5438]{Andrew Vanderburg}
\affiliation{Department of Physics and Kavli Institute for Astrophysics and Space Research, Massachusetts Institute of Technology, Cambridge, MA 02139, USA}

\author[0000-0002-5077-881X]{John F. Wu}
\affiliation{Space Telescope Science Institute, 3700 San Martin Drive, Baltimore, MD 21218, USA}
\affiliation{Department of Physics \& Astronomy, The Johns Hopkins University, Baltimore, MD 21218 USA}
\affiliation{Department of Computer Science, The Johns Hopkins University, Baltimore, MD 21218 USA}

\author[0000-0002-0579-6613]{M. Jimena Rodríguez}
\affiliation{Space Telescope Science Institute, 3700 San Martin Drive, Baltimore, MD 21218, USA}
\affiliation{Instituto de Astrofísica de La Plata, CONICET--UNLP, Paseo del Bosque S/N, B1900FWA La Plata, Argentina }

\author[0000-0002-4663-6827]{Remy Indebetouw}
\affiliation{University of Virginia Astronomy Department, P.O. Box 400325, Charlottesville, VA, 22904, USA}
\affiliation{National Radio Astronomy Observatory, 520 Edgemont Rd, Charlottesville, VA 22903, USA}

\author[0000-0003-0946-6176]{M\'ed\'eric Boquien}
\affiliation{Université Côte d'Azur, Observatoire de la Côte d'Azur, CNRS, Laboratoire Lagrange, 06000, Nice, France}

\author[0000-0002-0560-3172]{Ralf S. Klessen}
\affiliation{Universit\"{a}t Heidelberg, Zentrum f\"{u}r Astronomie, Institut f\"{u}r Theoretische Astrophysik, Albert-Ueberle-Str 2, D-69120 Heidelberg, Germany}
\affiliation{Universit\"{a}t Heidelberg, Interdisziplin\"{a}res Zentrum f\"{u}r Wissenschaftliches Rechnen, Im Neuenheimer Feld 225, 69120 Heidelberg, Germany}

\author[0009-0009-9167-3932]{Sophia Rivera}
\affiliation{Department of Physics \& Astronomy, The Johns Hopkins University, Baltimore, MD 21218 USA}

\author[0000-0002-5204-2259]{Erik Rosolowsky}
\affiliation{Dept. of Physics, University of Alberta, 4-183 CCIS, Edmonton, Alberta, T6G 2E1, Canada}

\author[0000-0001-9852-9954]{Oleg~Y.~Gnedin}
\affiliation{Department of Astronomy, University of Michigan, Ann Arbor, MI 48109, USA}

\author[0000-0002-5782-9093]{Daniel A. Dale}
\affiliation{Department of Physics \& Astronomy, University of Wyoming, Laramie, WY 82071, USA}

\author[0000-0003-3917-6460]{Kirsten~L.~Larson}
\affiliation{AURA for the European Space Agency (ESA), Space Telescope Science Institute, 3700 San Martin Drive, Baltimore, MD 21218, USA}

\author[0000-0002-8528-7340]{David A. Thilker}
\affiliation{Department of Physics and Astronomy, The Johns Hopkins University, Baltimore, MD 21218, USA}

\author[0000-0002-5259-2314]{Gagandeep Anand}
\affiliation{Space Telescope Science Institute, 3700 San Martin Drive, Baltimore, MD 21218, USA}

\begin{abstract}
The environments around star clusters evolve as stellar feedback reshapes the interstellar medium and dynamical processes reorganize the structure of the surrounding stellar field. As approximately single-age populations, star clusters can serve as clocks to trace these environmental changes. In this exploratory study, we test whether convolutional neural networks (CNNs) can identify age-dependent changes in cluster environments.  We take cluster ages as given from basic SED fitting of five-band UV–optical aperture photometry from the PHANGS (Physics at High Angular resolution in Nearby GalaxieS) HST survey. We first show that CNNs can be trained on image cutouts centered on clusters to recover ages directly from imaging. This demonstration provides the foundation for this study, which examines whether the information used by CNNs to predict age is coherent and physically meaningful. We perform controlled image occlusion experiments as an explainable AI method. These show that the CNNs extract age-predictive environmental cues in the absence of cluster light and when information on SED shape is removed by combining the five filters into one image. We find that reliance on environmental information increases at the youngest ($<$10 Myr) and oldest ($>$1 Gyr) ages, where clusters can exhibit similarly red colors. Our results are consistent with the long-recognized picture that cluster environments evolve systematically with age. We demonstrate that this information is encoded at a level detectable by machine-learning and recoverable from broadband imaging. This establishes a path for using new techniques to connect image-based age inference to the physical evolution of cluster environments.
\end{abstract}


\section{Introduction}

Stars form in dense environments, as members of associations or bound clusters, within clumps of gas resulting from the fragmentation of giant molecular clouds \citep[e.g.,][]{lada03, bp20}. These star clusters and associations represent fundamental units of the star formation process, where key aggregate properties, such as the initial mass function, multiplicity, and spatial distribution of stars, are imprinted \citep[e.g.,][]{lada03}. Their subsequent spatial and temporal evolution are governed by a combination of stellar feedback \citep[e.g.,][]{oort55,mckee77,bally16,sike25}, local gravitational dynamics, and galactic environmental conditions.  These processes determine the ultimate survival of the cluster or association, shape their properties, the state of the surrounding ISM, as well as the broader spatial distribution of stars in galactic disks \citep[e.g.,][]{meng22, farias24, Lahen25}.

As clusters age, their environments thus evolve, as residual gas and dust is dispersed, stellar densities decrease, and stars are ejected or drift from their birth sites. These changes are observable with high resolution data from facilities such as the Gaia, Hubble, and Webb space telescopes \citep[e.g.,][]{hannon19, brown21, dellacroce24, rodriguez25}. In principle, these spatial signatures can be linked back to a stellar population’s evolutionary stage, and it has been shown that environmental features, such as local crowding, the presence of nearby young stars, or the degree of hierarchical structure, are correlated with cluster age \citep[e.g.,][]{whitmore11, gouliermis15, krumholz19}.

The methods used to characterize this evolution generally capture selected, partial projections of the underlying physics. For example, measurements of cluster size, local density, or crowding rely on parametric measurements, which are then examined through distribution functions or two-dimensional correlations, for example through size-mass relationships \citep[e.g.,][]{brown21}, correlation function analyses \citep[e.g.,][]{grasha17, turner22}, and cluster mass functions \citep[e.g.,][]{larsen09, chandar10, cook23}. Such approaches have provided a solid observational basis for much of our understanding of astrophysics, and have done so by compressing the complexity of the underlying physical processes. 
However, the physics underlying star and cluster formation and evolution span many orders of magnitude, and it has been found that simultaneous observational and theoretical treatment across those scales are required to develop a coherent physical model.\citep[e.g.,][]{krumholz14}. This raises the possibility that essential physical relationships remain hidden from our standard reductive approach.

Convolutional neural networks (CNNs) enable a different means of analysis: rather than extracting localized, discrete measurements from image data, they operate directly on pixel-level data and can potentially identify multi-scale, non-linear patterns that could be difficult to otherwise extract. In this exploratory study, we exploit this feature to investigate how and where age-related information is encoded in the cluster environment. This differs from typical uses of CNNs -- i.e., object recognition, classification or label prediction \citep[e.g.,][]{Alexnet, wei20} -- and instead focuses on identifying spatial cues that the models draw on.

Specifically, we use CNNs to quantify the degree to which age can be predicted from different segments of imaging data, even without access to the cluster itself, using UV–optical broadband imaging from the PHANGS (Physics at High Angular Resolution in Nearby GalaxieS) HST survey \citep{lee22}, and the corresponding star cluster catalog \citep{maschmann24}. These data provide a well-characterized sample of star clusters and associations, with SED-based age estimates based on five-band aperture photometry \citep{turner21}. 
Our strategy is as follows.  We first demonstrate that CNNs are able to recover cluster age directly from HST image cutouts, when trained on the PHANGS SED-based ages. To investigate what information the CNNs rely on, we perform a series of image occlusion experiments as a form of explainable AI analysis \citep{xu2019explainable, dwivedi2023explainable}. In one set of experiments, we progressively mask the central pixels dominated by cluster light to test how age prediction performance changes as more of the cluster is removed. In the complementary “outer masking” experiments, we retain the cluster core while masking the surrounding field. We then collapse the five filters into a single composite image, to remove the color information on which the PHANGS cluster ages are based, and repeat these experiments.  This forces the CNNs to extract the predictive role of spatial structure and morphology alone. Taken together, these experiments provide a quantitative assessment of how age-correlated information is distributed between the clusters themselves, their colors, and their environments. 

The outline of the paper is as follows: Section~\ref{sec:Data} gives an overview of the PHANGS--HST cluster sample and describes the preparation of the images for model training.  Section~\ref{sec:Methods} describes the CNN architecture and provides the requisite demonstration that CNNs can be trained to recover cluster age directly from imaging. Section~\ref{sec:overview} then provides an overview of the series of experiments used to interrogate the models.  Section~\ref{sec:Results} presents the main findings from the occlusion and masking analyses, and Section~\ref{sec:Discussion} interprets these results in the context of cluster evolution, and discusses open issues that should be addressed in future work. Finally, we summarize the key conclusions in Section~\ref{sec:Conclusion}.

\section{Data Curation}
\label{sec:Data}

\subsection{PHANGS-HST Star Clusters}
The PHANGS-HST survey \citep{lee22} has produced the largest catalog to-date of $\sim$100,000 star clusters and associations across 38 nearby spiral galaxies \citep{maschmann24}.  For this exploratory analysis, we use a subset of visually inspected (i.e., human verified) star clusters and associations in 15 galaxies for which cluster ages were available at the time of analysis.  This subset spans the full range of distances in the PHANGS-HST galaxy sample (Table~\ref{tab:sample}) and provides $\sim$9000 sources for analysis.\footnote{Catalogs of the observed properties as described in \citet{maschmann24} are available on the PHANGS High Level Science Product (HLSP) website at MAST (\url{https://archive.stsci.edu/hlsp/phangs}. Catalogs of the physical properties, including age, have not yet been released at the time of the submission of this paper.}

The PHANGS star cluster catalog is V-band selected, and is the result of pipeline development efforts summarized in \cite{lee22}, which employed refined techniques for cluster candidate detection and selection \citep{whitmore21, thilker22}, aperture corrections \citep{deger22}, and automated morphological cluster classification using machine learning techniques \citep{wei20, whitmore21, hannon23}.  The survey itself, described in detail in \cite{lee22}, obtained 5-band NUV, U, B, V, I HST imaging with exposure times of $\sim$2200~s, $\sim$1100~s, $\sim$1100~s, $\sim$670~s, and $\sim$830~s, respectively.  Exposure times varied depending on whether imaging of sufficient depth and coverage was available in the archive; exact values for each galaxy are provided in \citet[][Table 1]{maschmann24}.  Survey observations were conducted with the HST WFC3/UVIS camera and the F275W, F336W, F438W, F555W, F814W filters (NUV, U, B, V, and I, respectively). In cases where HST ACS/WFC archival imaging of sufficient depth was available, the existing F435W data were used, and no new data were taken with WFC3 F438W.  The images were drizzled with a pixel scale of 0\farcs04; the size of the WFC3 V-band PSF is 0\farcs067.  Photometry was measured in circular apertures with radii of 4 pixels.  The corresponding physical extent of this aperture spans from $\sim$4--15 pc, as listed for each galaxy in Table~\ref{tab:sample}.  This aperture captures roughly $\sim$50--70\% of the light from the source, as measured from growth curves of reference samples of bright, relatively isolated clusters identified in each galaxy for the purposes of determining aperture corrections \citep{deger22}.

The PHANGS-HST catalog includes star clusters and associations spanning from $\sim$1 Myr to old globular clusters that formed a few hundreds Myr after the Big Bang. The sample covers more than four decades in stellar mass, up to $\sim10^{6}M_\odot$ \citep[][Table 1]{maschmann24}. The catalog provides a range of observed and physical properties for each cluster. In this study, we primarily use the source position (V-band DOLPHOT peak), the distance of the parent galaxy, and the age. 

The ages, along with stellar mass and dust reddening, were computed with CIGALE \citep{boquien19} which performs spectral energy distribution (SED) fitting.  The five-band HST NUV, U, B, V, and I aperture photometry for each source are provided as input to CIGALE, which derives the stellar mass by fitting the observed SED with single-age stellar population synthesis models and determining the normalization required for the model to match the observed luminosities. This normalization sets the total stellar mass, given the assumed initial mass function, so the mass is effectively constrained by the overall brightness of the cluster once its age and attenuation are determined from the colors.  

For the reference ages supplied to the CNNs, we adopt the initial set of PHANGS-HST cluster ages derived with a basic, standard method described in \citet{turner21}.  Based on tests with mock catalogues, \citet{turner21} estimate uncertainties of 0.3 dex in age, 0.2 dex in mass, and 0.1 mag in reddening (i.e., standard deviation of the difference between the model and recovered ages). Age uncertainties are largest at 1 Myr, around 10 Myr, and at the oldest ages, reflecting known degeneracies in SSP models at 5–50 Myr, while comparisons with earlier LEGUS results for NGC~3351 \citep{adamo17} show no significant systematic offsets. By using the initial PHANGS–HST age estimates derived solely from photometry \citep{turner21}, rather than later versions refined with environmental information \citep{thilker25}, we avoid circular reasoning and ensure that the CNN independently tests whether environmental features contain predictive information about age.

\begin{table*}[!htbp]
\centering
\begin{tabular}{lccccc}
\hline
Galaxy&D$[Mpc]$& $\sigma$(D)$[Mpc]$&ap size $[pc]$ &cutout size $[pc]$&N \\
(1)&(2)&(3)&(4)&(5)&(6)\\
\hline 
NGC~5068	&	5.20	&	0.21	&	4.0	&	113	&	326	\\
IC~5332	&	9.01	&	0.41	&	7.0	&	196	&	377	\\
NGC~0628c	&	9.84	&	0.63	&	7.6	&	214	&	530	\\
NGC~3351	&	9.96	&	0.33	&	7.7	&	216	&	116	\\
NGC~3627	&	11.32	&	0.48	&	8.8	&	246	&	958	\\
NGC~2835	&	12.22	&	0.94	&	9.5	&	265	&	893	\\
NGC~4254	&	13.10	&	2.8	&	10.2	&	285	&	747	\\
NGC~5248	&	14.87	&	1.34	&	11.5	&	323	&	593	\\
NGC~4321	&	15.21	&	0.49	&	11.8	&	330	&	950	\\
NGC~4535	&	15.77	&	0.37	&	12.2	&	343	&	531	\\
NGC~1566	&	17.69	&	2	&	13.7	&	384	&	850	\\
NGC~1433	&	18.63	&	0.56	&	14.5	&	405	&	252	\\
NGC~7496	&	18.72	&	2.8	&	14.5	&	407	&	321	\\
NGC~1512	&	18.83	&	0.88	&	14.6	&	409	&	424	\\
NGC~1365	&	19.57	&	0.78	&	15.2	&	425	&	783	\\
\end{tabular}
\caption{Sample of PHANGS-HST galaxies, star clusters and associations used in this analysis.  A total of 8651 stellar clusters and associations are included. The columns provide (1) the galaxy name, (2, 3) galaxy distance and error (based on PHANGS-HST TRGB measurements and compiled from the literature as published in \citealt{anand20, lee22}), (4) physical scale corresponding to the radius of the aperture used for PHANGS-HST cluster catalog photometry (4 pixels or 0\farcs16), on which PHANGS-HST SED-fit ages and photometry are based \citep{lee22, deger22} (5) physical scale corresponding to the size of the postage stamps provided to the CNNs (112 pixels), (6) number of human-classified star clusters and associations in the galaxy with coverage in all five imaging filters (NUV, U, B, V. I). Catalogs and data are available at \url{https://archive.stsci.edu/hlsp/phangs}.}
\label{tab:sample}
\end{table*}

\subsection{Image Cutouts and Normalization}
\begin{figure*}[!htbp]
\includegraphics[width=\textwidth]{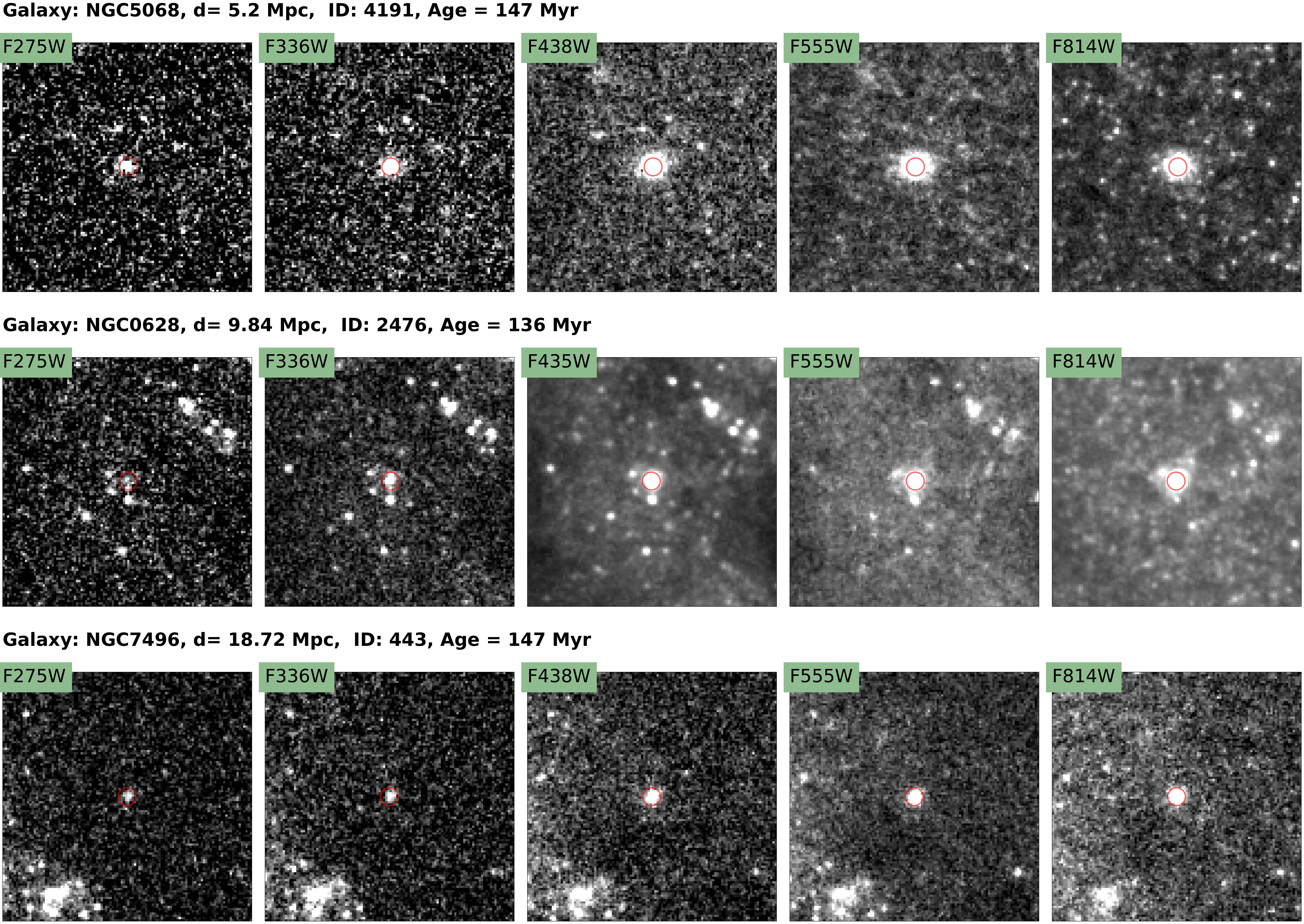}
\caption{Examples of the 112$\times$112 pixel HST image cutouts used in the CNN analysis.  Three star clusters are shown, each centered in the cutouts, with the full set of five band NUV U B V I images.  The examples are chosen to span the distance range of the PHANGS galaxy sample.  The red circle marks the adopted photometric aperture (radius of 4 pixels or 0\farcs14).  After applying a correction for light lost outside the aperture \citep{deger22}, this photometry is used to derive the cluster ages for the PHANGS cluster catalog through spectral energy distribution fitting as described in \citet{turner21}.  These cluster ages are the ones used to train the CNNs in this work.} 
\label{fig:cutouts}
\end{figure*}

For each star cluster, we produced image cutouts centered on the cluster position for each of the five filters. Clusters which did not have coverage in all five filters were dropped from the sample (e.g., at the edges of the observation footprint).  Each cutout was defined to be $112 \times 112$ pixels in size (subtending 4\farcs48), corresponding to a physical scale spanning $\sim$100--400 pc for galaxies at distances 5--20 Mpc (Table~\ref{tab:sample}).  In Figure~\ref{fig:cutouts} we show examples of the image cutouts for three clusters with ages $\sim$100 Myr that span the distances of the parent galaxies in the sample.
We initially generated five-band image cutouts for each cluster, and stored them as multi-extension FITS files. For efficient data loading, these were concatenated into a single HDF5 file, which serves as the input to the CNNs.

Normalization of the input data is a key component of our modeling approach, as pixel intensities vary substantially across HST filters due to both intrinsic cluster properties and background levels of the local environment within the parent galaxy. Without normalization, these variations can lead the CNN to focus on spurious differences in absolute pixel values rather than learning meaningful patterns in morphology, color, or environment \citep{sola1997importance}. Importantly, this normalization removes global differences in observed flux between clusters due to the distance of the parent galaxy. 

We evaluated several input normalization strategies, including per-pixel, per-channel (i.e., filter), and per-instance (i.e., cluster) approaches based on the pixel statistics of individual image cutouts. For our final experiments, we adopted a per-instance normalization scheme in which all pixel values within a $112 \times 112 \times 5$ image cube corresponding to a single cluster were scaled to the [0, 1] range using a single minimum and maximum computed across all five channels. This method preserves inter-channel flux ratios, which are essential for capturing astrophysically meaningful color information.

We therefore adopt per-instance, cross-filter normalization as the standard for all models trained with five-filter input. For experiments using mean-combined single ``white-light" images, the same per-instance min–max scaling was applied.

\subsection{Data Augmentation}

To improve model generalization and enforce rotational invariance, we applied data augmentation by rotating each input cutout by $90^\circ$, $180^\circ$, and $270^\circ$, increasing the training set size by a factor of four. Because cluster orientation on the detector is arbitrary and carries no physical meaning, this augmentation prevents the CNN from learning detector-dependent correlations and encourages it to focus on physically relevant morphological features.

Indeed, we also tested additional augmentations via horizontal and vertical flipping, which would have expanded the dataset by a factor of eight in total. However, we found that the inclusion of flipped images provided no improvement in predictive performance beyond the augmentation of the dataset from rotations.
For the final experiments reported in this work, we retained only the fourfold augmentation based on rotations.

For all experiments, the dataset was then divided into training, validation, and testing subsets with a standard 70\%--15\%--15\% ratio, corresponding respectively to optimization of the model weights, monitoring of performance during training and hyperparameter tuning, and an independent benchmark of predictive generalization.

\section{Methods}
\label{sec:Methods}

\subsection{CNN Architecture}

\begin{table*}[t]
\centering
\caption{Summary of the adopted CNN architecture.}
\label{tab:table_cnn_architecture}
\begin{tabular}{lccccc}
\hline
\textbf{Layer Type} & \textbf{Filters/Units} & \textbf{Kernel/Pool} & \textbf{Stride} & \textbf{Activation} & \textbf{Output Shape} \\
\hline
Input            & ---  & ---       & --- & ---   & $112\times112\times5$ or $112\times112\times1$ \\
Conv2D           & 128  & $3\times3$ & 1   & ReLU  & $110\times110\times128$ \\
MaxPooling2D     & ---  & $2\times2$ & 2   & ---   & $55\times55\times128$ \\
Conv2D           & 256  & $3\times3$ & 1   & ReLU  & $53\times53\times256$ \\
MaxPooling2D     & ---  & $2\times2$ & 2   & ---   & $26\times26\times256$ \\
Conv2D           & 512  & $3\times3$ & 1   & ReLU  & $24\times24\times512$ \\
MaxPooling2D     & ---  & $2\times2$ & 2   & ---   & $12\times12\times512$ \\
Flatten          & ---  & ---        & --- & ---   & $73{,}728$ \\
Dense            & 512  & ---        & --- & ReLU  & $512$ \\
Dense            & 256  & ---        & --- & ReLU  & $256$ \\
Dense            & 128  & ---        & --- & ReLU  & $128$ \\
Dense (Output)   & 1    & ---        & --- & Linear& $1$ \\
\hline
\end{tabular}
\end{table*}

We adopted a convolutional neural network (CNN) framework to model the mapping between star cluster imaging data and cluster ages. CNNs are well-suited to this task given their ability to capture spatial patterns and morphological features in multi-channel images \citep[e.g.,][]{Lecun98,Alexnet,zf2013}, and their demonstrated success in a range of astrophysical applications \citep{2015MNRAS.450.1441D,2015ApJS..221....8H,2018MNRAS.473.3895L,WuBoada19,2020ApJ...900..142W,2021ApJ...914..142H,2022ApJ...927..121W,2023ApJ...954..149D,2023PASA...40....1H,Graham2025}, including for the classification of star cluster morphologies in the PHANGS-HST cluster catalog pipeline \citep[e.g.,][]{wei20, hannon23}.

The CNN is trained to predict the PHANGS-derived logarithmic cluster age, $\log(\mathrm{age/yr})$, as a continuous regression target, using the image cutouts described in the previous section and illustrated in Figure~\ref{fig:cutouts}.

The adopted CNN architecture (Table~\ref{tab:table_cnn_architecture}) consists of three convolutional blocks, each followed by a max-pooling layer to progressively reduce spatial dimensions while refining the feature encoding used by the network to capture relevant spatial structure. The convolutional layers employ ReLU activation functions \citep{Nair2010} to introduce non-linearity and enable complex feature extraction from the five-channel input images. Thus, the output of the convolutional stack is flattened and passed through three fully connected layers of progressively decreasing size, before a single linear output neuron predicts the logarithmic cluster age.

The model was trained using the Adam optimizer \citep{kingma2017adammethodstochasticoptimization} with an initial learning rate of $1\times10^{-6}$ and a mean squared error (MSE) loss function (i.e., goodness-of-fit metric used to validate performance of the model) based on the predicted $\log(\mathrm{age})$ and the target reference values. 

All the hyperparameters were selected empirically based on validation performance, with MSE on $\log(\mathrm{age})$ as the criterion.  We chose a batch size of 32 (the number of training examples the network processes at once before updating its weights) and trained for up to 200 epochs (a complete pass through the entire training dataset).  We iteratively varied learning rate (i.e. step size), batch size, and epochs, and chose the set of hyperparameters that minimized validation error while remaining stable and avoiding overfitting.

We also employed an \texttt{EarlyStopping} callback \citep{prechelt2002early} (patience = 30) to prevent overfitting when the validation loss plateaued (i.e., 30 consecutive epochs with no improvement in goodneess-of-fit are allowed). All the models were implemented in \texttt{TensorFlow} \citep{abadi2016tensorflow} and trained on MIT’s Engaging supercomputer equipped with 2 NVIDIA A100 GPUs and 128 CPU cores, allocating 100GB of RAM per run.

On average, to train a set of 5 runs for one case required about 20 minutes with this setup. For reference, to produce Figure~\ref{fig:blackout_experiments} (40 cases per plot, 4 plots, and 5 models per case) required training $40 \times 4 \times 20$ minutes in total, i.e. $\sim$53 hours of compute time.

For all experiments (as we describe in Section~\ref{sec:occlusion}), we adopted the same CNN architecture and hyperparameter configuration.  This consistency allowed for a fairer comparative analysis across cases. In some experiments, early stopping terminated training well before the maximum number of epochs, mitigating concerns about overestimating the epoch limit. In every case, we verified from the loss curves that overfitting did not occur and that the validation loss reached a plateau, indicating no apparent undertraining. We therefore consider the results presented here to be directly comparable across cases. We note, however, that a more thorough investigation would involve independently optimizing the architecture and hyperparameters for each case, a process that would require a substantial computational effort. Nonetheless, in a few isolated cases where alternative architectures were explored, the resulting performance was very similar to that obtained with the chosen configuration.

As with any machine-learning analysis, the quantitative results will exhibit some dependence on the adopted architecture, optimization settings, and hyperparameters. We chose a conservative configuration designed to remain fixed across all experiments so that the ranging-radius and masking tests could be compared on equal footing, without re-tuning the model for each case. While alternative architectures or training strategies may yield slightly different numerical performance, we find that the qualitative trends reported here (particularly the spatial distribution of age-predictive information) are stable across reasonable choices. We therefore consider the results to be representative and robust for the purposes of this exploratory study.

\begin{figure*}[!htbp]
\includegraphics[width=\textwidth]{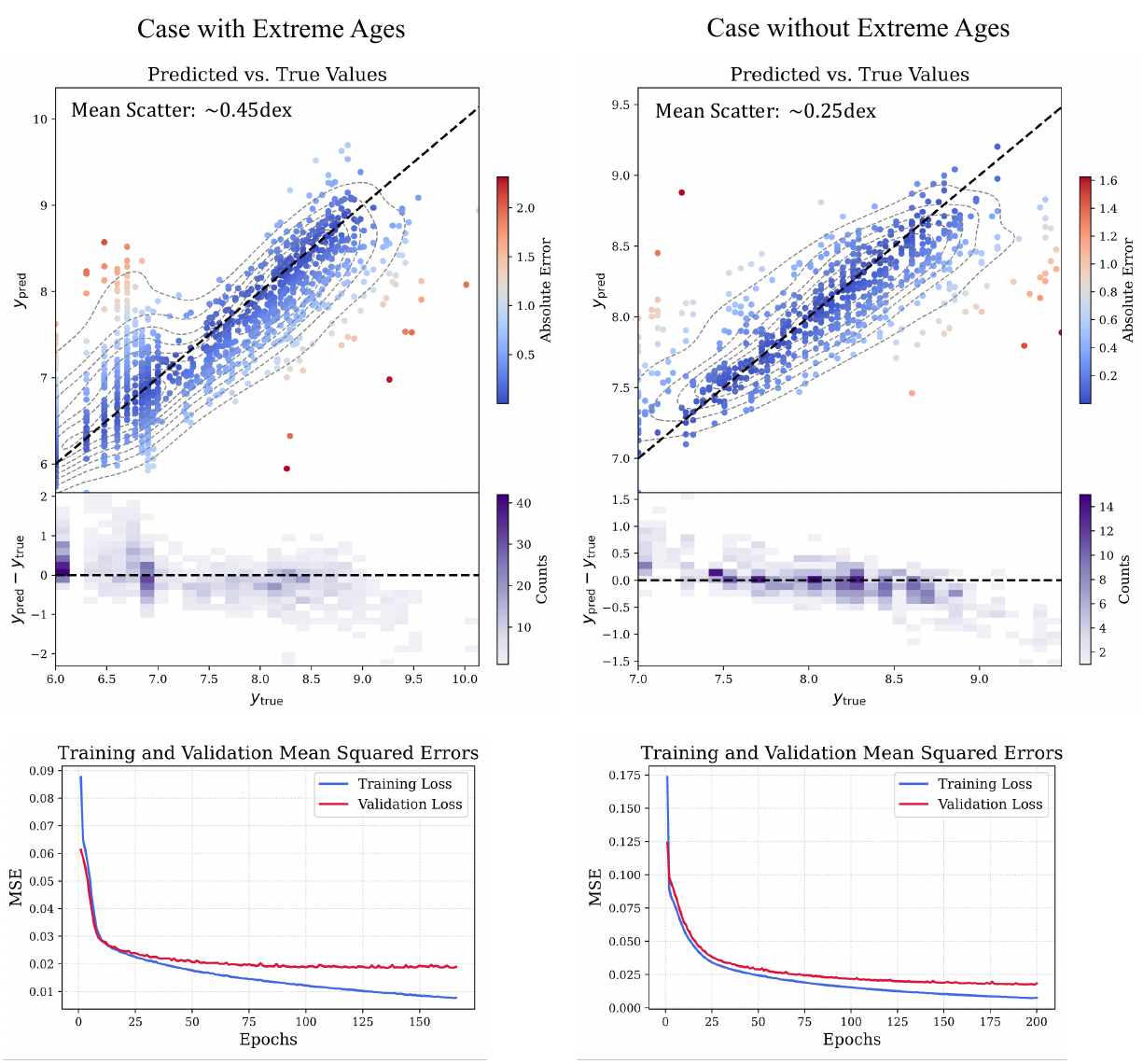}
\caption{CNN age-prediction performance for two training samples: (left) training and evaluation on the full PHANGS-HST cluster sample and (right) training and evaluation with the youngest and oldest clusters removed. The top panels show predicted versus reference ages, demonstrating that the CNN can indeed recover the SED-fit ages from the imaging but exhibits systematic residuals at the extreme ages. The middle panels plot residuals as a function of true age, highlighting the overestimation of very young clusters and the underestimation of very old clusters in the full-sample model. The bottom panels display training and validation mean-squared error (MSE) as a function of the number of passes through the training set. Together, these comparisons confirm the feasibility of age recovery from imaging.}
\label{fig:true_vs_pred}
\end{figure*}

\subsection{Establishing Analysis Feasibility: CNN Age Prediction from Five-Band Imaging}
\label{sec:feasibility}

Underlying the feasibility of this analysis is the assumption that CNNs can recover the reference PHANGS-HST ages directly from the imaging. We thus first demonstrate this capability, keeping in mind that our goal is not to develop a method for improving age estimates per se, but to use the resulting models to probe which spatial and environmental features carry age-related information. 

However, it is well known that the reliability of the reference ages is not uniform across the full parameter space. The youngest clusters ($\lesssim$ 10 Myr) and the oldest clusters ($\gtrsim$ 3 Gyr) suffer from well-known degeneracies in SED-based age estimates: both can appear comparably red, either because old populations are intrinsically red or because young populations can be reddened by dust or affected by sparse sampling of massive stars \citep[e.g.][]{krumholz15, hannon19, whitmore20, Henny2025}. Yet, these same clusters are also the ones for which we might expect the strongest age-related signatures in the surrounding environment.  We therefore evaluate model performance both with and without the extreme-age clusters to examine these issues.  Specifically, we train CNNs on (1) the full cluster sample and (2) a subset excluding clusters younger than 10 Myr ($\log(\mathrm{age/yr}) < 7$) or older than 3 Gyr ($\log(\mathrm{age/yr}) > 9.5$).

Our results for CNN image based recovery of ages derived via SED modeling are presented in Figure~\ref{fig:true_vs_pred}.  Using a held-out test set, the CNN model trained on the 5-filter cutouts (with per-instance normalization penalizing reliance on the absolute flux, as discussed above) achieves a median scatter relative to the reference SED-fit ages of 0.45 dex in the the full sample, and of 0.25 dex in the sample without the youngest and oldest clusters (Figure~\ref{fig:true_vs_pred}, top panels).

Residuals ($y_\mathrm{pred} - y_\mathrm{true}$) are shown in the middle panels of Figure~\ref{fig:true_vs_pred}. These are centered near zero across much of the age range, but the spread narrows in the intermediate regime ($7.5 < \log(\mathrm{age/yr}) < 9.0$) and increases toward the boundaries as might be expected. This pattern indicates that the CNN is most reliable for intermediate-age clusters, while recovery of the reference cluster ages become less stable for the youngest and oldest objects. The reduction in scatter when excluding the youngest and oldest clusters is consistent with the expected impact of degeneracies in SED-based age estimates at these extremes. This behavior supports the interpretation that the CNN is recovering physically meaningful trends from the imaging, and that the remaining performance limitations may partly reflect uncertainties in the reference ages rather than an absence of age-predictive signal in the data. On the other hand, the sign of these residuals reflects a tendency to overestimate the ages of very young clusters and underestimate those of very old clusters. This systematic behavior is consistent with known degeneracies in broadband SED-based age estimates, and also partly reflect a statistical regression-to-the-mean effect in the model predictions, where extreme values in the training labels are preferentially drawn toward the central range of the age distribution. Finally, training and validation loss curves (Figure~\ref{fig:true_vs_pred} bottom panels) converge smoothly without signs of overfitting, confirming that the models are well-optimized under both training sets (i.e., for the full sample, and without the extreme-age clusters).

\section{Overview: Disentangling Sources of Predictive Information}
\label{sec:overview}
To determine which information the CNNs use when predicting cluster ages, we perform a series of controlled experiments that vary both the input images and the composition of the training and testing datasets. These include occlusion studies that mask different parts of the images, as well as experiments in which the five photometric bands are collapsed into a single stacked image.

Because the reference PHANGS ages are derived from five-band SED fitting, retaining the full multi-channel input would allow the CNNs to recover age directly from implicit color information present in the images. Collapsing the filters suppresses this pathway by removing broadband SED information, thereby preventing the models from inferring age through photometric color.

In this configuration, any remaining predictive performance must arise from spatial structure alone, forcing the CNNs to rely on cluster morphology and environmental context. The resulting stacked white-light images therefore provide a controlled test of whether age-related information is encoded in the spatial distribution of light rather than in the SED itself.

As a control, we compute the performance of a null predictor that does not use the input images at all. This predictor assigns the same constant value (the mean cluster age from the training set) to every input, so its residuals are simply the differences between the reference ages and this fixed mean. While not a model in any meaningful sense, it provides a lower bound for comparison: any effective CNN should outperform the naive strategy of always returning the mean. Note that this baseline is more conservative than random guessing, which would result in even higher errors for the predictions.

Together, the five-filter baseline (from the initial feasibility analysis to demonstrate recovery of the PHANGS reference ages from the HST image cutouts, Section~\ref{sec:feasibility}) and the null predictor define the upper and lower bounds of achievable age-prediction performance of our experiment. The controlled tests that follow characterize the predictive power that lies between these limits.  In this section, we provide an brief overview of the experiments performed. Both this overview and the compact summaries in Figures \ref{fig:five_cases} and \ref{fig:xtrms_vs_no_xtrms} later in the paper are intended to help orient the reader and provide a roadmap for the remainder of the paper.

\subsection{Use of 5-Filter vs. Mean-Combined Filter Images}

As already discussed, we train CNNs on five-channel inputs (Section~\ref{sec:feasibility}), with each channel corresponding to one of the HST filters. This setup preserves the full color information as well as morphological detail.

To remove color as a variable, we also trained models on mean-combined images (a stacked white-light image), in which the five filters are averaged to produce a single-channel input to the CNN. The images are first averaged and then normalized, in order to preserve the signal-to-noise weighting of each filter. In this representation, spectral energy distribution (SED) information (i.e., color) is suppressed, preventing the CNN from trivially recovering age from the broadband SED and instead forcing it to rely on spatial structure. This allows us to probe the spatially encoded age-predictive features learned by the CNN. 

\subsection{Image Occlusion Experiments}
\label{sec:occlusion}
To identify the spatial structures that are predictive of age, and to examine how this information is distributed across spatial scales within the image cutouts, we performed a series of occlusion experiments in which circular regions of progressively increasing size were masked. In one set of experiments, a circular mask of increasing diameter (0–80 pixels) was applied at the cluster center, progressively removing the core and then the surrounding regions (inside-out masking). In a complementary set, the inverse mask was applied, progressively removing the outer regions while retaining the central pixels (outside-in masking). Together, these experiments allow us to measure the model’s sensitivity to age-predictive features as a function of radial distance from the cluster center.

Masked pixels were set to zero and kept fixed at that value; they were not scaled nor included in the computation of the normalization. Normalization was then applied to the remaining pixels based on their non-zero values, while preserving the original $112 \times 112$ input dimensions. Experiments were carried out for both the five-filter and stacked white light inputs to test the dependence on color information. Each configuration was repeated with five independently trained models, and performance was quantified using the mean logarithmic scatter on held-out test sets (clusters excluded from training and used only for evaluation). 

\section{Results}
\label{sec:Results}


\subsection{Radial Distribution of Age-Predictive Information}
\label{sec:innermask}

\begin{figure*}[!htbp]
\includegraphics[width=\textwidth]{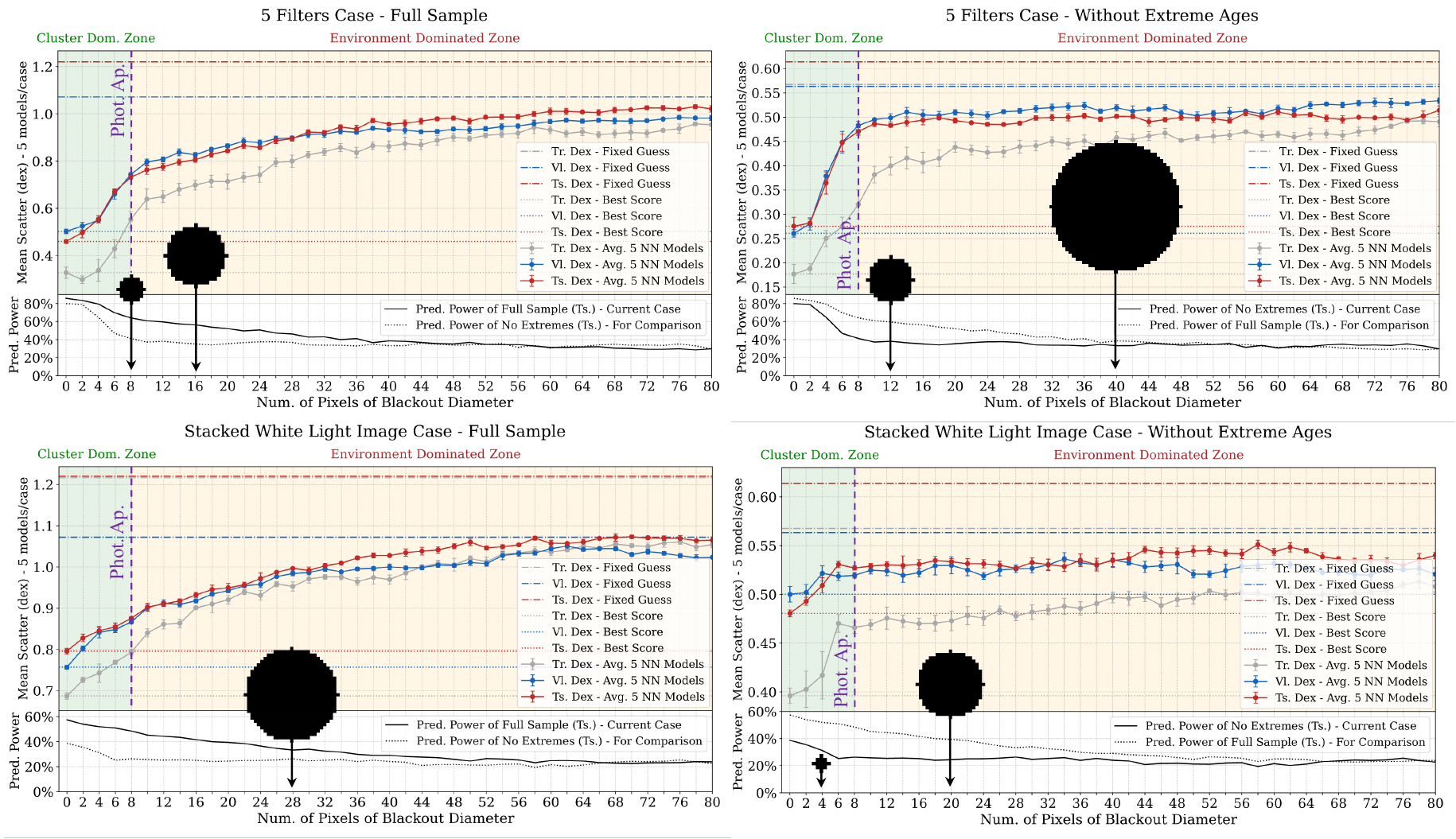}
\caption{
    Blackout experiments to assess the spatial localization of age-predictive information in image cutouts centered on star clusters.  A circular mask of increasing diameter is applied concentrically over each image, progressively removing central and surrounding regions. The $x$-axis shows the blackout diameter (in pixels), while the $y$-axis reports the mean scatter (dex) of five independently trained models for training (Tr.), validation (Vl.), and test (Ts.) datasets. The top row corresponds to the 5-filter HST input, and the bottom row to the stacked single white-light image. Left panels use the full cluster sample, while right panels exclude extreme-age systems. Dashed horizontal lines at the top of each panel indicate the fixed-guess baseline (non-learned average prediction). The black silhouettes illustrate representative blackout masks at different diameters. The different color regions mark the transition from cluster-dominated zones (green) to environment-dominated zones (orange). The black curve at the bottom of each panel quantifies the relative predictive power retained in the visible pixels, computed $(\mathrm{baseline}_{\mathrm{test}}^{2} - \mathrm{curve}_{\mathrm{test}}^{2}) / \mathrm{baseline}_{\mathrm{test}}^{2}$. These experiments quantitatively demonstrate that age-predictive information is present in the environment surrounding clusters.}
    
\label{fig:blackout_experiments}
\end{figure*}

Figure~\ref{fig:blackout_experiments} presents results from four experiments in which the central regions of the image cutouts are progressively masked.

We show both the five-filter and the single stacked white-light cases (top and bottom panels, respectively), each evaluated for the full cluster sample (left panels) and for a subsample excluding the youngest and oldest clusters (right panels).  In each panel, two vertically stacked plots are shown with masking diameter on the shared x-axis.  In the upper plots, the curves trace the mean logarithmic scatter from five independently trained models as a function of the masking diameter, shown separately for the training, validation, and test sets. The lower plots show the relative predictive power remaining in the visible pixels for the test set, to facilitate comparison of results from the four experiments. Since adjacent pixels are not statistically independent due to the instrumental PSF, these experiments probe spatially correlated morphological structures rather than independent pixel-level measurements. The curves in the lower plots are calculated as
\[
\frac{\mathrm{baseline}_{\mathrm{test}}^{2} - \mathrm{curve}_{\mathrm{test}}^{2}}{\mathrm{baseline}_{\mathrm{test}}^{2}}.
\]


The figure reveals clear trends with the size of the masked region. In both five-filter experiments (top panels of Figure~\ref{fig:blackout_experiments}), the error rises most steeply as the mask diameter increases from 0 to $\sim$8 pixels, indicating that the central region of the cluster carries the strongest age-predictive signal. Notably, this scale matches the 8-pixel diameter circular aperture used for the photometry underlying the reference ages (corresponding to 4--15 pc in radius from the cluster center for the distances of the galaxies in this analysis as given in Table~\ref{tab:sample}), suggesting that the CNN has implicitly learned this aperture.

In contrast, the single-image experiments (bottom two panels of Figure~\ref{fig:blackout_experiments}) show a less pronounced break in the curve, implying that the connection to the 8-pixel region arises primarily from color information. This is consistent with expectations, since color is the principal driver of the SED-fit ages.

Beyond a diameter of 8 pixels, model performance continues to decline in all cases, although more gradually.  Importantly, they always remain below the null predictor, showing that the CNN models do continue to extract signal even when much of the cluster is hidden. The predictive-power curves at the bottom of each panel quantify this residual information as a fraction of the null predictor, expressed in dimensionless percentage error to facilitate comparison across experiments. Together, these diagnostics quantitatively demonstrate that while the reference ages are tied to the photometric aperture (the half-light radius), the CNN can not only learn this scale but also can identify additional age-predictive signal in the outer regions of the cluster and its environment -- information that was not used to derive the reference ages. It has long been recognized that the age of a cluster and the appearance of its surrounding environment are closely linked; these results show that this relationship can be quantitatively studied with CNNs from the broadband imaging.

Further insight comes from comparing the full-sample panels (left column) with those excluding the extreme-age clusters (right column). Removing the youngest and oldest systems reduces the overall scatter by about half in both the five-filter and single white-light experiments (note the different y-axis ranges). However, although the full-sample cases exhibit higher absolute error, their predictive-power curves remain higher than those of the restricted set. This contrast indicates that the environments of the youngest and oldest clusters provide especially valuable age-predictive information. Consistent with this, the curves for the full sample continue to rise beyond the 8-pixel photometric aperture, whereas those for the restricted sample flatten. The sharper degradation in the full sample demonstrates that masking the environment disproportionately harms predictions when extreme-age clusters are included, underscoring the essential role of environmental context in reliably constraining the ages of the youngest and oldest systems.

\subsection{Separating the Roles of Color, Cluster Light, and Environment}
\label{sec:separating}

\begin{figure*}[!htbp]
\includegraphics[width=\textwidth]{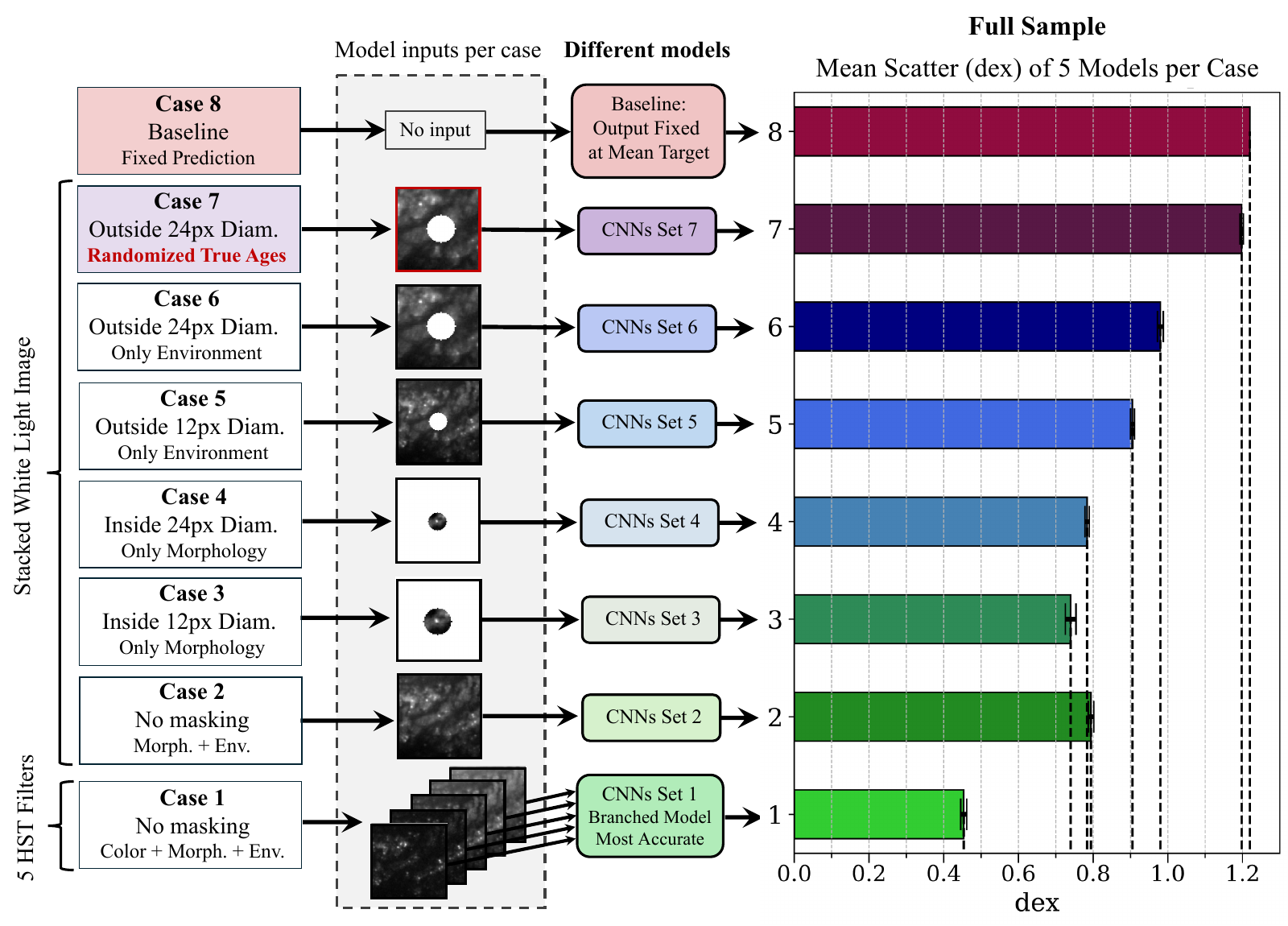}
\caption{Performance comparison of 8 cases for cluster age prediction, using the full range of cluster ages (log(age) from 6.0 to 10.0). Case 1 includes unmasked cutouts in all 5 HST filters and provides the maximum information on cluster color, morphology, and environment, as reported in Section~\ref{sec:feasibility} and Figure~\ref{fig:true_vs_pred}. Case 2 uses a stacked white-light image, retaining only morphology and environment. Cases 3 and 4 use the stacked image with the environment removed using 24-pixel and 12-pixel diameter outer masks, respectively, thereby focusing the model on the cluster light. Cases 5 and 6 apply the mask inverse, masking out the cluster center with 12-pixel and 24-pixel diameters, respectively, thereby focusing the model on the environment, as reported in Section~\ref{sec:innermask} and Figure~\ref{fig:blackout_experiments}. Case 7 uses the same outside-24-pixel environment input as Case 6, but with randomized reference ages during training and validation. This prevents the network from learning any physically meaningful correlations, causing it to converge to predictions near the mean age of the training set and to exhibit performance comparable to the null baseline. Case 8 shows the performance of the null predictor, providing a lower bound for comparison. The bars show the mean scatter (dex) of log(age), which is adopted as the performance metric. Each case is based on 5 independently trained CNN models, with error bars showing variability across models. All results refer to the test set.}
\label{fig:five_cases}
\end{figure*}

\begin{figure*}[!htbp]
\includegraphics[width=\textwidth]{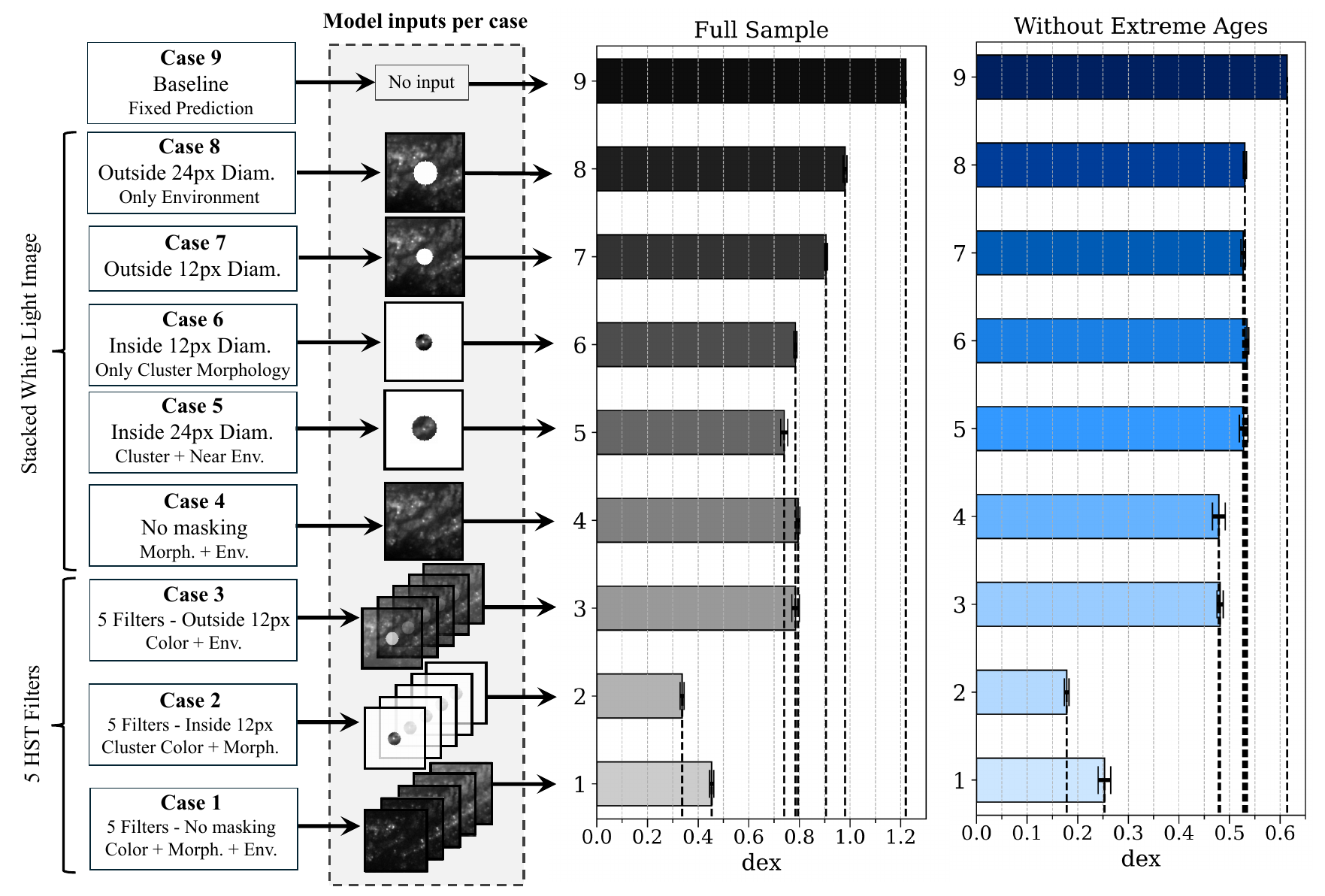}
\caption{Similar to Figure~\ref{fig:five_cases}, but now also showing results for models trained on a restricted set of reference ages (log(age) = 7.0–9.5), in which the youngest and oldest clusters have been excluded (right panel of bar charts). Cases 1 and 4–8 are reproduced from Figure~\ref{fig:five_cases} in the left panel of bar charts. Two additional occlusion experiments using the full five-filter input are added (Cases 2 and 3 in this figure), applying outer and inner masking, respectively.}
\label{fig:xtrms_vs_no_xtrms}
\end{figure*}

To complement the radius-dependent occlusion results obtained by progressively masking the images from the cluster center outward (inside-out masking; Figure~\ref{fig:blackout_experiments}), we perform a broader set of experiments that also include masking regions outside the cluster (outside-in masking). The results of these experiments are summarized in Figures~\ref{fig:five_cases} and \ref{fig:xtrms_vs_no_xtrms}. These experiments include eight and nine cases, respectively, designed to separate the roles of color, cluster light, and environment.

Instead of tracking performance as a function of masking radius, these figures summarize the test‐set scatter for each case side by side. 
Figure~\ref{fig:five_cases} reports results for the full cluster sample, while Figure~\ref{fig:xtrms_vs_no_xtrms} shows the corresponding results with the youngest and oldest clusters removed. For reference, we include the expected scatter from a null predictor that assigns a fixed mean age to all clusters, shown as the top case in each figure. Improvements relative to this worst-case scenario reflect the amount of predictive information in each case.

Figure~\ref{fig:five_cases} shows results using the complete range of cluster ages (log(age) = 6.0–10.0) for training. Case~1 refers to the analysis using the full image cutouts (no masking) in all five HST filters, and the resulting scatter is 0.45~dex, as reported in Section~\ref{sec:feasibility}.  As expected, this model yields the lowest scatter among the cases considered in this Figure, since it uses the maximum amount of multi-filter data.  Case~2 collapses the five filters into a single stacked white-light image, yielding a scatter of $\sim$0.8~dex. As expected, performance is worse than Case~1, since color information is removed. Nevertheless, although the loss of color reduces predictive power, the model retains enough signal from cluster morphology and environment to perform significantly better than the null predictor baseline ($\sim$1.2 dex). 

Cases~3 and~4 focus the models on the cluster-only light by masking out the environment outside 24-pixel and 12-pixel diameter apertures, respectively. Because an 8-pixel diameter aperture contains roughly half the cluster’s light \citep{deger22}, Case~4 captures mainly the central morphology, whereas Case~3 also includes the extended cluster emission that reaches out to $\sim$20–24 pixels. Case~3 performs slightly better (0.74~dex vs.\ $\sim$0.8~dex for Case~4), but both are comparable to Case~2. Together, these results show that excluding the environment beyond $\sim$12–24 pixels produces only a modest degradation in performance, with both masked cases yielding scatter comparable to the white-light image. This implies that, in the absence of color information, the dominant predictive signal is confined to the cluster and its immediate surroundings.

Cases~5 and~6 apply the mask inverse, which now removes the cluster from the white light image, and forces the model to only use information in the environment. In Case~5, the central 12 pixels are masked, yielding a scatter of $\sim$0.9 dex. In Case~6, the mask extends to 24 pixels, fully excluding the cluster and retaining only the surrounding galactic environment, with a scatter of $\sim$1.0 dex. Although the overall scatters are large, both cases still perform better than the baseline ($\sim$1.2 dex), indicating that the environment alone does carry some age-predictive signal.

Case~7 serves as another control test: it uses the same environment-only input as Case~6, but with randomized age labels during training and validation. As expected, the model is unable to learn meaningful correlations and converges to predictions close to the mean-age baseline. The fact that performance in Case~7 is indistinguishable from the null predictor demonstrates that the gains in Case~6 are not an artifact of model architecture or training procedure, but instead arise from genuine age-predictive information encoded in the environment. 

These experiments confirm, in a complementary way to the central masking analysis as a function of radius, that age information is not confined to the cluster light itself. 

In summary, cluster-scale structure provides most of the age-predictive signal in the absence of color information. If both color information is removed and the cluster is masked, the surrounding environment still carries a measurable, albeit weak, predictive signal.

Next, we examine how these trends change when the youngest and oldest clusters are removed (Figure~\ref{fig:xtrms_vs_no_xtrms}).

Figure~\ref{fig:xtrms_vs_no_xtrms} extends this analysis by comparing nine cases for both the full sample of reference ages (bar charts in left panel) and a restricted set excluding the youngest and oldest clusters (bar charts in right panel). We add the results from an additional two cases where the masking experiments have been performed on the 5-filter cutouts (Cases 2-3).  In the full-sample experiments, Case~9 provides the fixed-mean predictor baseline with a scatter of $\sim$1.2~dex, while in the restricted-age experiments the corresponding baseline is lower at $\sim$0.6~dex. Each set of models are judged by their relative improvement over these baselines.

In the age-restricted sample, Cases~7 and~8 (which mask out the cluster and retain only the environment outside 12- and 24-pixel radii) achieve accuracy comparable to the cluster-only Cases~5 and~6, all with scatters of $\sim$0.54 dex. This indicates that, once extreme-age clusters are removed, the surrounding environment provides only limited additional predictive information, carrying roughly the same weight as the cluster alone. In other words, for clusters at intermediate ages, environmental cues contribute less to age prediction relative to the dominant role of color.

Another noteworthy result is shown in Case~2, which focuses on the cluster light within 12 pixels and its colors across all five filters. In the age-restricted sample, Case~2 achieves the best overall performance at $\sim$0.18 dex, compared to $\sim$0.25~dex for the full five-filter cutouts (Case~1) and $\sim$0.6~dex for the baseline. In the full sample of cluster ages, the same pattern holds: Case~2 reaches $\sim$0.33~dex, better than Case~1 at $\sim$0.45~dex and far better than the $\sim$1.2~dex baseline. Despite discarding environmental information, Case~2 narrows the inputs to a more focused and relevant region, effectively sharpening the model’s attention. This illustrates a general effect in our experiments: removing less-informative regions can improve performance by forcing the model to concentrate on the most predictive features, in a manner conceptually similar to attention mechanisms in transformer architectures.

Taken together, these results reveal a nuanced dependence on both age regime and the choice of image inputs into the model. Excluding extreme ages simplifies the task and lowers the scatter, but it also removes much of the additional leverage provided by environmental context. 

This outcome mirrors the decision-making process used by observers when checking for egregious errors in the SED-fit cluster ages by eye \citep[e.g.,][]{hannon22}. Observers rely on the surrounding environment to distinguish degenerate extreme-age solutions, but for clusters most likely at an intermediate age, color information alone is usually sufficient. The CNN analysis independently recovers this same logic, demonstrating its potential to formalize and quantify established empirical practice in a consistent, automated way.


\subsection{Dependence on Galaxy/Cluster Distance}

\begin{figure*}[!htbp]
    \centering
    \includegraphics[width=\textwidth]{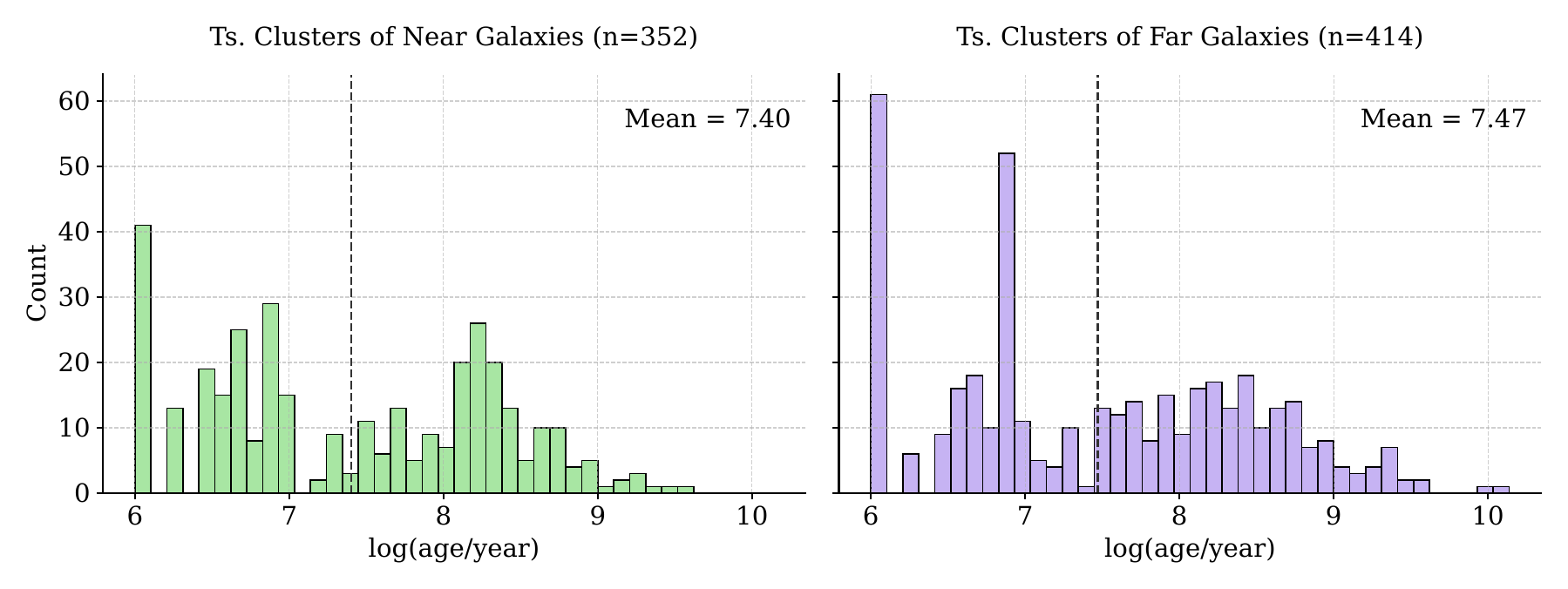}
    \caption{
    Distribution of cluster ages for two parent galaxy samples, divided by galaxy distance, the nearest between 5-12 Mpc (IC~5332, NGC~0628c, NGC~3351, NGC~3627) and the furthest between 17-20 Mpc (NGC~1566, NGC~1433, NGC~7496, NGC~1512, NGC~1365). Although the mean ages are very similar (7.40 vs. 7.47), the distant galaxy sample contains a large fraction of very young clusters around log(age/year) $\approx 6$, where the age determination is most degenerate, as well as some of the oldest clusters in the dataset. By contrast, the nearest set of galaxies has more clusters at intermediate ages. These differences in age distribution complicate the interpretation of Figure~\ref{fig:blackout_near_far}, since it is not straightforward to determine in which regime the model error should be expected to dominate.
    }
    \label{fig:hist_near_far}
\end{figure*}

\begin{figure*}[!htbp]
    \centering
    \includegraphics[width=\textwidth]{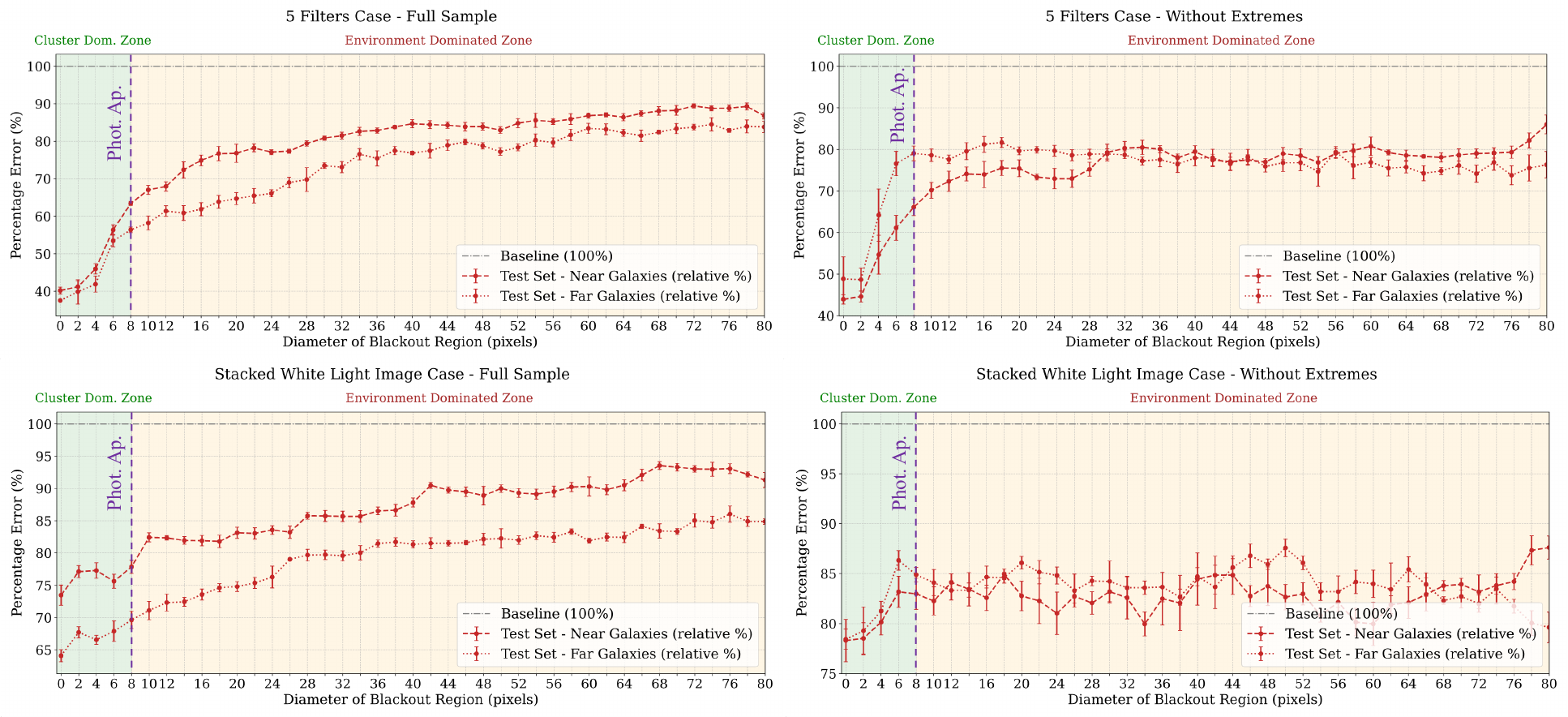}
    \caption{
   Occlusion experiments with inner-cluster masking, analogous to Figure~\ref{fig:blackout_experiments}, but separately for the nearest (IC~5332, NGC~0628c, NGC~3351, NGC~3627) and most distant (NGC~1566, NGC~1433, NGC~7496, NGC~1512, NGC~1365) galaxy samples. The top row shows results for the five-filter inputs, and the bottom row for the stacked white-light images. The left panels use the full dataset, while the right exclude extreme-age clusters. The x-axis gives the blackout diameter (in pixels), and the y-axis shows relative percentage error. This metric is computed by dividing the scatter (dex) from each blackout configuration by the scatter of the corresponding fixed-guess baseline (mean cluster age of the training set), then expressing it as a percentage. The dashed black line marks the 100
    }
    \label{fig:blackout_near_far}
\end{figure*}

Our analysis shows that color, cluster light, and the surrounding environment each contribute age-predictive information detectable by a CNN; however, the results may depend on spatial resolution. To test this, we repeated the occlusion analysis with two sub-samples of clusters in the nearest (between 5-12 Mpc: IC~5332, NGC~0628c, NGC~3351, NGC~3627) and furthest galaxies (between 17-20 Mpc: NGC~1566, NGC~1433, NGC~7496, NGC~1512, NGC~1365), and present the results in Figures~\ref{fig:hist_near_far} and ~\ref{fig:blackout_near_far}.

 In Figures~\ref{fig:blackout_near_far}, both the multi-filter (color, top panels) and single-filter (white-light, bottom panels) cases, the same qualitative behavior is recovered: errors rise steeply as the central pixels are masked, followed by a slower but persistent increase as the masked area expands outward. Thus, the basic conclusion of the masking analysis, that both the cluster and its environment carry age-predictive information, does hold across galaxies spanning a range of distances.

The detailed behavior of the near- (dotted) and far-galaxy (dashed) curves reflects a combination of effects. 
Clusters in nearby systems subtend a larger area on the image and portions should of the cluster, in principle, remain detectable even when masked.
Yet the results for the nearest galaxy subsample tends to have larger errors, particularly when the full range of cluster ages is considered.  Differences in the age distributions of the two sub-samples (Figure~\ref{fig:hist_near_far}), are significantly driven by the fact that more actively star-forming and rarer galaxies are preferentially found at larger distances.  This complicates the comparison, and degeneracies at the youngest and oldest ages further bias the error curves. For these reasons, the detailed results are difficult to interpret, but both subsamples nonetheless reproduce the same overall trends. This demonstrates that the conclusions of this study are robust against distance-dependent effects.

\section{Discussion}
\label{sec:Discussion}

\subsection{Open Issues and Physical Interpretation}
This study presents the first application of a convolutional neural network to extract physical insight on the environments of star clusters from imaging. Consequently, several limitations and open questions remain, and the following points of uncertainty should be kept in mind.

\begin{enumerate}
 
    \item \textit{Origins of scatter in age predictions}: 
    The scatter observed when training on the full age range likely reflects a combination of effects (Figure~\ref{fig:true_vs_pred}), including errors in the SED-derived reference ages (due to age-metalicity-reddening degeneracies, as well as degeneracies resulting from stochastic sampling of the stellar initial mass function for low mass clusters), distance and resolution-dependent biases, and potentially suboptimal model tuning. It is unclear the extent to which the observed scatter is driven by limitations in the reference age estimates rather than by the CNN models themselves.

    \item \textit{Robustness of relative performance trends between ``cases"}: 
    Cases~2--4 in Figure~\ref{fig:five_cases} yield comparably large scatters ($\sim$0.74--0.8~dex), and the substantial overlap in their error bars warrants caution in over-interpreting small differences in performance. While Figure~\ref{fig:outer_masking} provides additional evidence for a negative correlation between outer masking radius and model performance, the statistical significance of these differences remains uncertain.

    \item \textit{Choice of angular versus physical resolution}: 
    Our experiments adopt image cutouts and masking defined at a constant \textit{angular} resolution, allowing the physical scale per pixel to vary with distance. This approach ensures that the CNN can optimally learn phenomena at the highest resolution available.  Nonetheless, adopting a constant \textit{physical} resolution represents an important avenue for future work. Such tests are non-trivial, however, as the models would no longer be exposed to the same number or distribution of non-zero input pixels, complicating direct performance comparisons.

    \item \textit{Potential ``information leakage'' from spatial overlap}: 
    Because young clusters are themselves clustered, their corresponding image cutouts may include partially overlapping regions. This raises the possibility that some pixels appear in both the training and test sets, introducing mild information leakage. While this effect is expected to be limited and unlikely to dominate the model behavior, it nonetheless represents a caveat of the current dataset construction. Future iterations of this experiment will incorporate stricter spatial partitioning to control this source of leakage.

\end{enumerate}

Taken together, these uncertainties complicate a detailed physical interpretation of the result presented in Figure 3-5. Nonetheless, it is instructive to consider, at a qualitative level, what physical information the network may be responding to.

Star formation proceeds hierarchically, with GMC-scale complexes fragmenting into associations and compact clusters \citep[e.g.,][]{elmegreen2000}. Young clusters often appear alongside nearby substructures, dust lanes, and embedded companions, whereas older clusters, particularly globular clusters, tend to reside in smoother fields after their natal gas and associated substructure have been dispersed. The CNN may therefore be exploiting signatures of this hierarchy: structured, multi-component environments for the youngest clusters and increasingly homogenized surroundings for the oldest. This interpretation is consistent with our long-standing understanding of clustered star formation and suggests that, even in the presence of imperfect reference ages, CNNs can extract quantitative information about the temporal evolution of cluster environments directly from broadband imaging.

\subsection{Future Work}

While the current analysis leaves a number of open questions, those questions provide a clear path for future studies which can improve both the physical interpretability and predictive power of CNN-based analyses of star cluster imaging.

\begin{enumerate}

    \item \textit{Improving reference ages}: 
    Future work to improve the adopted reference ages naturally divides into two categories. Of considerable interest is investigating whether unsupervised or semi-supervised machine-learning approaches can improve cluster age estimates, for example by identifying structures in the images that could help break degeneracies inherent in SED-based ages.
    On the other hand, improving the underlying reference ages via additional observations would enhance the robustness of explainable machine-learning techniques that link image-based age inference to the physical evolution of cluster environments. For example, additional photometry or spectroscopy could be added. \citet{Henny2025} and \citet{thilker25} have improved age estimates by combining SED-based ages with environmental priors.  However, using environment-informed age estimates during training would complicate subsequent attempts to interpret the CNN as learning environmental signatures, and care must be taken to avoid circularity.
    
    \item \textit{CNN model interpretability}: 
    More sophisticated interpretability techniques, such as gradient-based saliency maps, integrated gradients, and feature-attribution methods \citep{saliency-mapping,integrated-gradients,GradCAM,2020ApJ...900..142W,Ntampaka+2022,2024ApJ...967..152A}, can be used to better understand which image regions and structures drive the CNN predictions. Interpretation of the neural network is also possible through sparse decomposition of network activations, or circuits that form the network \citep[e.g.,][]{k-sparse-autoencoders,saes,Wu2025,interpretable-sae-astro}.  While the radial masking experiments presented here provide a deliberately simple, coarse-grained form of such an analysis, these more advanced techniques would enable a finer, non-parameteric, pixel-level view of the features influencing the model output. In principle, this could allow us to distinguish the relative importance of the cluster core, extended stellar envelopes, nearby dust lanes, ISM structures, and surrounding star-forming substructure. Such analyses would help determine whether the CNN is responding to physically meaningful features consistent with expectations for cluster and environmental evolution, or to more subtle or unexpected image cues, and deepen our understanding of the physical interpretability of the results.

    \item \textit{Physical-scale masking}: 
    As noted in the previous section, future work could explore masking schemes defined in terms of physical (parsec-scale) radii, rather than fixed angular scales, in order to better normalize experiments across galaxies at different distances. Such an approach would enable more direct comparisons of the spatial scales contributing to the CNN predictions, but would require careful control of resolution- and sampling-related effects, because differences for example in input image size or pixel rescaling could cause the CNN to encode artifacts of the imaging pre-processing rather than physically meaningful structure.

    \item \textit{Input image normalization}: 
    Exploring alternative input normalization schemes represents another potential direction to improve model performance. Non-linear transformations, such as logarithmic scaling, may enhance the relative contribution of low-surface-brightness environmental features relative to the bright cluster core. Such approaches are common in astronomical imaging \citep{2004PASP..116..133L} and increasingly used in astrophysical machine-learning applications \citep{angeloudi2024the}, but remain relatively uncommon in conventional computer-vision pipelines. Quantile-based normalization of the input pixel values may also help mitigate variations in dynamic range and signal-to-noise across filters. Because the image backgrounds are dominated by highly structured astrophysical emission rather than blank sky or instrumental noise, different normalization choices will directly alter the relative weighting of cluster features and incidental line-of-sight structure. A systematic exploration of these effects is therefore warranted.

    \item \textit{What's the added value over human validation of ages?} 
    Future work can test how to apply CNN-based approaches for age validation. Relative to human visual inspection, ML-approaches can of course to scale consistently to tens of thousands of clusters, and potentially detect subtle, multi-scale environmental signals that may be difficult to identify by eye. Establishing the advantages (and disadvantages) more clearly will be important for positioning machine-learning methods as a tool to augment expert analysis.

\end{enumerate}

Taken together, these future directions are aimed at moving beyond demonstrating that age information is present in cluster environments toward a deeper understanding which aspects of that environment carry the signal and why. While the current analysis establishes that the immediate few-hundred-parsec surroundings of star clusters encode age-dependent information accessible to CNNs, it does not yet isolate the physical structures responsible for this dependence.

A particularly informative regime is likely to be clusters for which the CNN predictions and SED-based ages disagree. Rather than treating such cases as failures, they provide an opportunity to identify systematic patterns in the local environments that may not be fully captured by traditional age-dating methods. Human inspection of these discrepant cases may reveal qualitative differences -- such as whether a cluster resides in an extended association, a clumpy or smooth stellar field, or an environment that appears unusually young or evolved -- that can help guide more targeted, quantitative analyses. In an upcoming work, we will discuss example clusters where ML predictions and SED-fit ages disagree (J. Wu et al., \textit{in prep}). Some of these cases highlight failures of the ML model, while in other cases, we identify clusters that likely have SED fitting errors (e.g., due to aperture contamination or insufficient treatment of dust).

\subsection{Machine learning as a more widely adopted tool for cluster evolution?}

This work highlights a path for using machine learning in astrophysics that goes beyond treating it as a large-scale optimization exercise on ever-growing datasets or collections of human-labeled images. Gaining new scientific insight requires more than simply improving predictive performance; it requires understanding what information the models use and why it is physically relevant. 

Expert human interpretation illustrates this point: much of what astronomers infer during image inspection (e.g., whether to verify ages as in \citealt{hannon22} and \citealt{whitmore23} or classify cluster morphology as in \citealt{adamo17} and \citealt{wei20}) reflects a synthesis of intuition, accumulated experience (and bias), and physical understanding. This knowledge is difficult to formalize precisely because it depends on interpreting clusters within their complex local environments and against the often messy background of their host galaxies.

In this study, we turn that challenge into a strength by allowing models to operate directly on the multiscale imaging context, rather than treating the environment as background to be removed. The goal is not to replace expert judgment, but to establish an iterative workflow in which models surface both expected and unexpected, non-parametric patterns, and humans assess whether they have physical meaning or instead reflect artifacts of the modeling. 
Achieving this requires interpretable models and deliberate human steering. Simply interrogating an computational black box is not intellectually satisfying; developing approaches that make model behavior understandable -- and that keep the process genuinely engaging and meaningful for human researchers -- 
will be essential for overcoming skepticism and enabling machine learning to become a widely adopted and trusted tool for advancing discovery in astrophysics \citep[also see recent discussion in][]{2026arXiv260210181H}.

\section{Conclusions}
\label{sec:Conclusion}

We present an exploratory convolutional neural network (CNN) analysis of PHANGS–HST imaging of 15 nearby spiral galaxies, containing over 8000 star clusters and compact associations, to investigate where and how age information is encoded in broadband UV–optical images.  Using cluster ages based on SED modeling of 5-band aperture photometry as reference ``labels,'' we trained CNNs on multi-filter image cutouts centered on clusters, and systematically modified the input images through controlled masking and filter stacking (channel reduction) experiments. These experiments allowed us to disentangle the relative contributions of color, cluster light, and the surrounding environment to age prediction. Rather than aiming to improve age estimates themselves, our goal was to use machine learning as a diagnostic tool to investigate the spatial distribution of age-related information in and around star clusters. The main conclusions of this study can be summarized as follows:

\begin{enumerate}
    \item \textbf{We show that CNNs trained on image cutouts centered on star clusters which subtend $\sim$100-400 pc can recover cluster ages directly from the imaging}, achieving a scatter of $\sim$0.2–0.3 dex relative to reference ages derived from five-band UV–optical photometric SED fitting (Figure~\ref{fig:true_vs_pred}).
    
    \item \textbf{Age-predictive information is not confined to cluster light or color alone.} 
    Controlled masking and filter-stacking experiments demonstrate that CNNs retain age sensitivity even when the central cluster light is masked and when color information is removed by collapsing the five filters into a single white-light image (Figures~\ref{fig:five_cases}). This shows that age-related information is encoded in the surrounding environment as well as in the cluster itself.

    \item \textbf{Environmental information contributes measurably to age predictions.} 
    Progressively masking the environment surrounding the cluster leads to continued degradation of age-predictive power, indicating that spatial context beyond the cluster core contains non-negligible age information (Figures~\ref{fig:blackout_experiments}). This demonstrates that the environment plays an active role in the CNN predictions rather than serving merely as background.

    \item \textbf{Environmental cues are more important in age-degenerate regimes.} 
    The reliance on environmental information increases at the youngest ($\lesssim$10~Myr) and oldest ($\gtrsim$1~Gyr) ages, where broadband colors alone provide limited discrimination and clusters of very different ages can exhibit similar colors (Figure~\ref{fig:xtrms_vs_no_xtrms}). At intermediate ages, color information dominates the predictions.

    \item \textbf{The CNNs learn age-correlated signal in a self-consistent manner.} 
    The largest increase in scatter occurs when masking the central $\sim$0--8 pixels, corresponding to the photometric aperture used to derive the SED-based reference ages. This indicates that the models have learned the imprint of the labeling procedure itself, providing a reassuring internal consistency check rather than evidence of spurious correlations.
      
    \item \textbf{Results appear to be robust to distance effects.} Repeating the analysis for sub-samples of the nearest (5-12 Mpc) and furthest galaxies (17-20 Mpc) reproduces the same qualitative trends, including the strong sensitivity of age prediction to masking of the central cluster region and the continued degradation as larger spatial scales are removed (Figure~\ref{fig:blackout_near_far}). While the absolute scatter differs between subsamples, no single distance regime dominates the results.

\end{enumerate}

Taken together, these results are consistent with the long-recognized physical picture in which young clusters reside in structured, dusty, and hierarchically nested environments, while older clusters occupy progressively smoother fields as the imprint of star formation is erased. Specifically, the models exhibit increased sensitivity to environmental information when broadband colors are most degenerate( at very young and very old ages), and show continued degradation of age-predictive power as the environment is progressively masked.  Thus, the analysis demonstrates that age-dependent environmental information is quantitatively recoverable from broadband imaging at a level detectable by machine learning. At the same time, the radial masking experiments employed here provide only a simple, coarse-grained probe of this information: they establish that age-predictive signal resides on specific spatial scales, but do not by themselves isolate the particular morphological structures or physical features driving the CNN predictions.

Even with imperfect age labels, the CNNs extract coherent environmental signals, motivating a framework in which clusters act as clocks and their surrounding environments trace empirical evolutionary sequences across galactic scales. As a proof of concept, this work demonstrates the feasibility of using deep learning to connect resolved star cluster populations to the physics of star formation and feedback, while underscoring the need for interpretable CNNs to link model behavior to specific physical processes.

Looking ahead, this approach has the potential constrain the timescales of cluster dissolution or reveal how long the structure of the ISM retains memory of past star formation activity.  More broadly, it raises the possibility that environmental morphology itself could serve as an auxiliary age diagnostic, complementing traditional photometric or spectroscopic methods for unresolved clusters and helping to break longstanding degeneracies among age, reddening, and metallicity. Realizing this potential will require improved age calibrations, more careful treatment of resolution-dependent effects, and continued development of interpretable, physically grounded machine-learning methodologies.

\section{Data availability}
All the {\it HST} data used in this paper can be found in MAST: \citep{phangs_hst_data1, phangs_hst_data2}. The image cutouts and derived data products used in this analysis 
are also available on Zenodo: \dataset[10.5281/zenodo.18871344]{https://doi.org/10.5281/zenodo.18871344}.

\section{Code availability}
All code used to generate the results presented in this paper, along with representative trained models and outputs, is publicly available at \citep{Viana2025StarClustersCode}.

\vspace{5mm}
\facilities{HST(STIS), Swift(XRT and UVOT), AAVSO, CTIO:1.3m,
CTIO:1.5m,CXO}

\software{astropy \citep{2013A&A...558A..33A,2018AJ....156..123A},  
          Cloudy \citep{2013RMxAA..49..137F}, 
          Source Extractor \citep{1996A&AS..117..393B}
          }

\appendix

\section{Appendix: Outer Masking Experiments}
\label{sec:outer_masking_app}

\begin{figure*}[!htbp]
    \centering
    \includegraphics[width=0.75\textwidth]{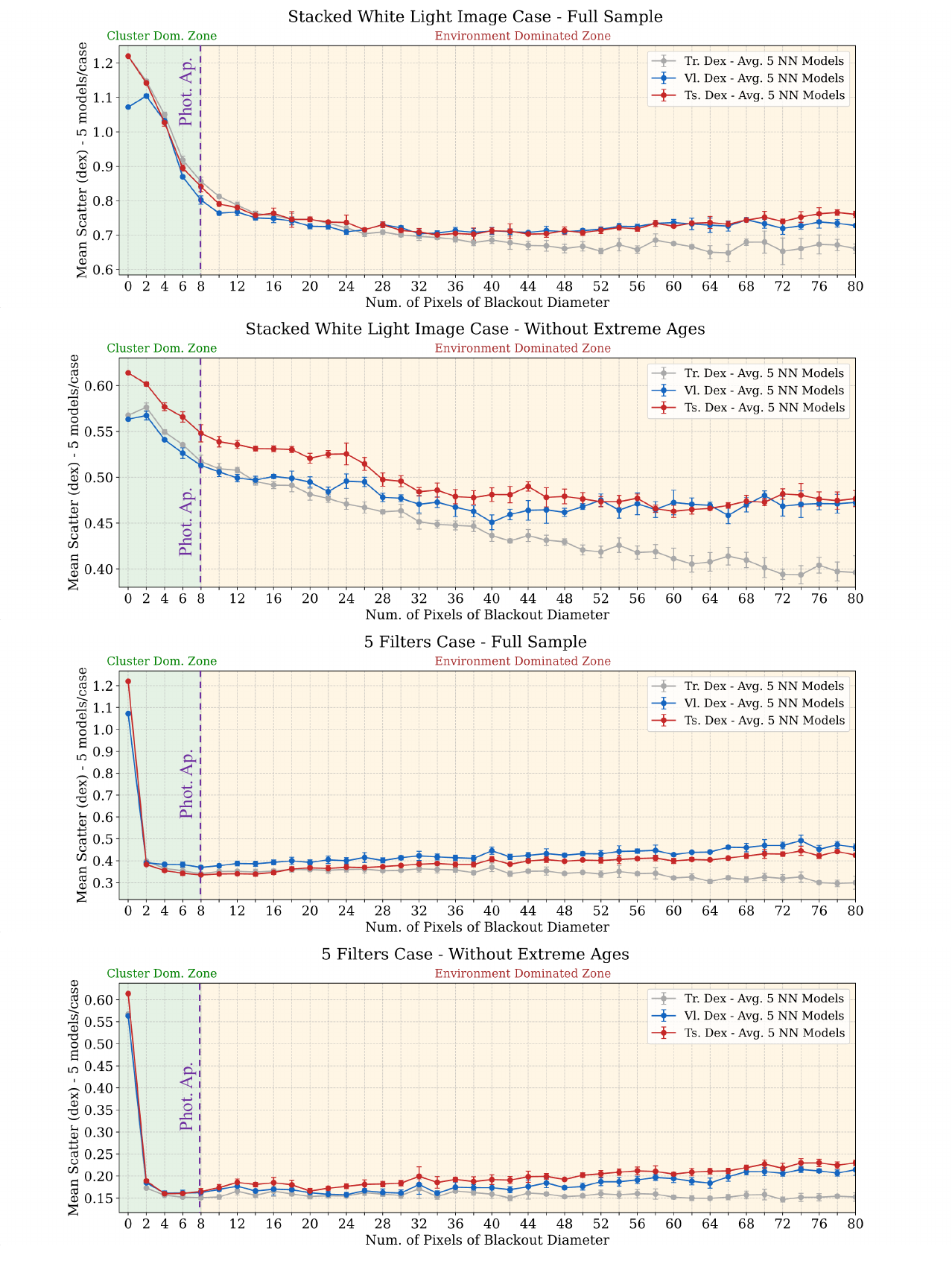}
    \caption{
    Results of the outer masking experiments. The plots show the mean test scatter as a function of the diameter of the visible region around each cluster, for the five filter case and the stacked image case, and for both the full sample and the sample without extreme ages. Dashed lines indicate the fixed guess baseline. These experiments provide the full set of complementary results to the inner masking analysis.
    }
    \label{fig:outer_masking}
\end{figure*}

For completeness, we show in Figure~\ref{fig:outer_masking} the results of the outer masking tests, the complement to the inner masking experiments shown in the main text Figure~\ref{fig:blackout_experiments}. These runs preserve the central cluster region while progressively masking the surrounding field, allowing us to quantify how predictive performance changes as environmental information is removed.

Two competing effects are visible in these curves. First, enlarging the visible diameter increases the dimensionality of the input, making the learning task harder. This is most evident in the five filter case, where each radial step adds five new rings instead of one, producing a rise in scatter driven purely by the larger input volume. Second, as the model gains access to more of the surrounding field, the error also decreases because the environment contains age–relevant information. The observed trends therefore reflect the balance between these two effects: increased difficulty from higher dimensional inputs, and improved performance from the additional environmental signal.

The lower panels of Figure~\ref{fig:outer_masking} show that the minimum scatter occurs at a diameter of about 8 pixels. This result is unsurprising, and occurs by design: our CNN is optimized to estimate SED modeled cluster ages \textit{based solely on five-band photometry captured in a radius of 4 pixels}. There is no additional information that \textit{can} be captured from cluster environments, because the \textit{ground truth} ages do not know about photometric information outside of the aperture. If instead, we had used spectroscopy or other information (beyond the NUV-U-B-V-I photometry) to define our cluster age ground truth, then it would be possible to test whether the CNN leverages extended image features to learn this alternate definition of cluster age. However, in our current experimental set up using the SED-fit ages, we are unable to evaluate whether there is additional \textit{color} information in cluster environments; instead, in these lower panels, we only observe that the CNN performance suffers with increasing number of unmasked pixels.

\section*{Acknowledgments}

Based on PHANGS (Physics at High Angular resolution in Nearby Galaxies) observations made with the NASA/ESA Hubble Space Telescope, obtained from MAST (Mikulski Archive for Space Telescopes) at the Space Telescope Science Institute. STScI is operated by the Association of Universities for Research in Astronomy, Inc. under NASA contract NAS 5-26555.  Support for Program number 15654 was provided through a grant from the STScI under NASA contract NAS5-26555.

JV acknowledges support from the Mauricio and Carlota Botton Foundation. MB acknowledges support by the ANID BASAL project FB210003. This work was supported by the French government through the France 2030 investment plan managed by the National Research Agency (ANR), as part of the Initiative of Excellence of Université Côte d’Azur under reference No. ANR-15-IDEX-01. MB also acknowledges support from the French National Research Agency (ANR), grant ANR-24-CE92-0044 (project STARCLUSTERS).  
RSK acknowledges financial support from the ERC via Synergy Grant ``ECOGAL'' (project ID 855130), from the German Excellence Strategy via the Heidelberg Cluster ``STRUCTURES'' (EXC 2181 - 390900948), and from the German Ministry for Economic Affairs and Climate Action in project ``MAINN'' (funding ID 50OO2206). 
We thank the German Science Foundation (DFG) for financial support in the project STARCLUSTERS (funding ID KL 1358/22-1 and SCHI 536/13-1).
ER acknowledges the support of the Natural Sciences and Engineering Research Council of Canada (NSERC), funding reference number RGPIN-2022-03499, and the Canadian Space Agency, reference number 23JWGO2A07.

\bibliography{main}{}
\bibliographystyle{aasjournal}

\end{document}